\documentclass[aps,prl,groupedaddress,showpacs,preprint]{revtex4}

\usepackage{subfigure}
\usepackage{graphics}
\usepackage{bm}

\newcommand{\bra}[1]{\langle #1|}
\newcommand{\ket}[1]{|#1\rangle}
\newcommand{\braket}[2]{\langle #1|#2\rangle}

\begin{document}

\title{Interstitial Electronic Localization}

\author{Bruno Rousseau}
\email[]{br75@cornell.edu}
\author{N.W. Ashcroft}
\affiliation{Laboratory of Atomic and Solid State Physics, Cornell University, Ithaca,
 New York 14853-2501}
\date{\today}

\begin{abstract}
We investigate the ground-state properties of a collection of \textit{N} non-interacting
electrons in a macroscopic volume $\Omega$ also containing a crystalline array of 
\textit{N} spheres of radius $r_c$ each taken as largely impenetrable to electrons and
with proximity of neighboring excluding regions playing a key physical role. The sole 
parameter of this quantum system is the ratio $r_c/r_s$, where $r_s$ is the Wigner-
Seitz radius. Two lattices (FCC and BCC) are selected to illustrate the behavior of the
system as a function of $r_c/r_s$. As this ratio increases valence electrons localize in
the interstitial regions and the relative band-width $\epsilon_F/\epsilon_F^0$ is found
to decrease monotonically for both. The system is motivated by the behavior of the
alkali metals at significant compression. It accounts for band narrowing, leads to
electronic densities with interstitially centered maxima, and can be taken as a model
which clearly may be improved upon by perturbation and other methods.
\end{abstract}

\pacs{71.23.An,71.20.Dg,71.10.Ca}

\maketitle

Wigner and Seitz \cite{Wigner1933}\cite{Wigner1934} originally observed that the major
physical properties of sodium at one atmosphere ($10^5$ Pa) could be accounted for by
considering the crystal as composed of two components, namely the ion cores,
constituted by the nuclei and their bound electrons, and itinerant valence electrons,
assumed statistically separate from the cores and in a paramagnetic state. For the
stabilizing volume they found that the cores exerted only a weak perturbation on the
valence electrons (leading to the nearly free electron (NFE) approximation) the core then
occupying a relatively small fraction of the unit cell. As a starting point for the theory
of the metallic state under ordinary conditions, the Wigner and Seitz description has
been very successful over the years. However, high pressure experiments conducted
recently are now challenging this point of view: when the cores are induced to occupy an
increasingly larger fraction of the unit cell the indications are that a new paradigm may
be appropriate, as is suggested here.

According to system, pressure can reduce linear dimensions by as much as 50\% and
under these conditions the alkalis depart notably from their expected "simple metal" (or
NFE) behaviors\cite{Maksimov2005}. Their crystal structures at room temperature
generally proceed from BCC to FCC, and then to non-close packed (for a concise
exposition of the structures adopted by the alkalis, see \cite{katzke184101} and
references therein). The latter are difficult to understand intuitively within the simple
NFE and \textit{nuclear-centric} viewpoint of Wigner and Seitz. Theoretical work also
indicates a breakdown of the NFE model: Neaton \textit{et al.} \cite{Neaton1999}
\cite{Neaton2001}, among others, reported \textit{ab initio} calculations for the band
structure of lithium and sodium at high compression. As they showed, the bands are far
from NFE-like and the occupied band-width is much smaller than the familiar
$50.1/r_s^2$ eV expected for free electrons. They also noted \cite{Neaton2001} that
the combined effects of Coulomb repulsion, Pauli exclusion, and orthogonality result in an
increase of valence electron density in the \textit{interstitial} regions the valence
electrons evidently being forced away from the near core regions characteristic of
nuclear-centric electron distributions.

\begin{figure}
\subfigure[ FCC ]{
\resizebox{80mm}{!}{\includegraphics{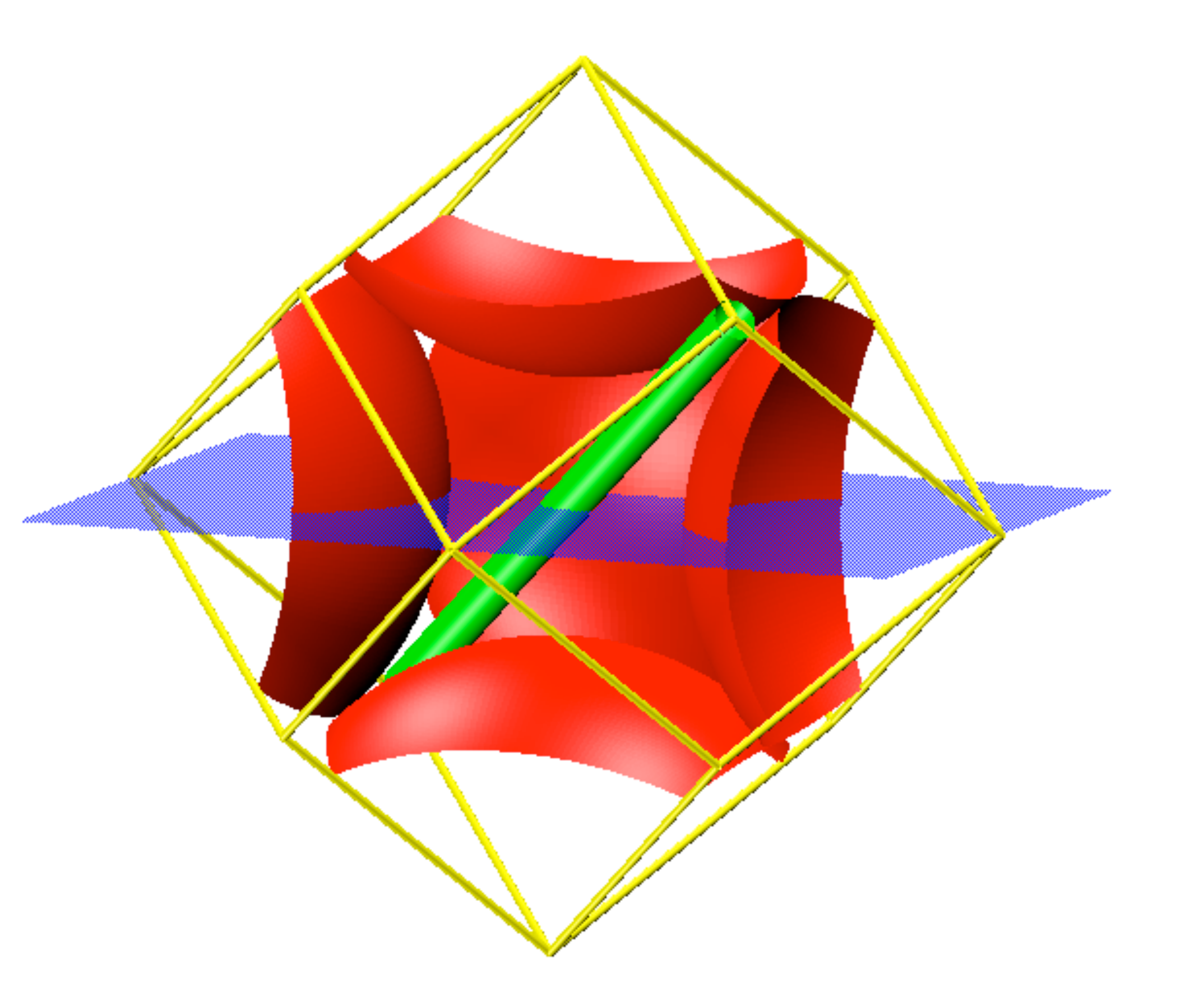}}}
\subfigure[ BCC]{
\resizebox{70mm}{!}{\includegraphics{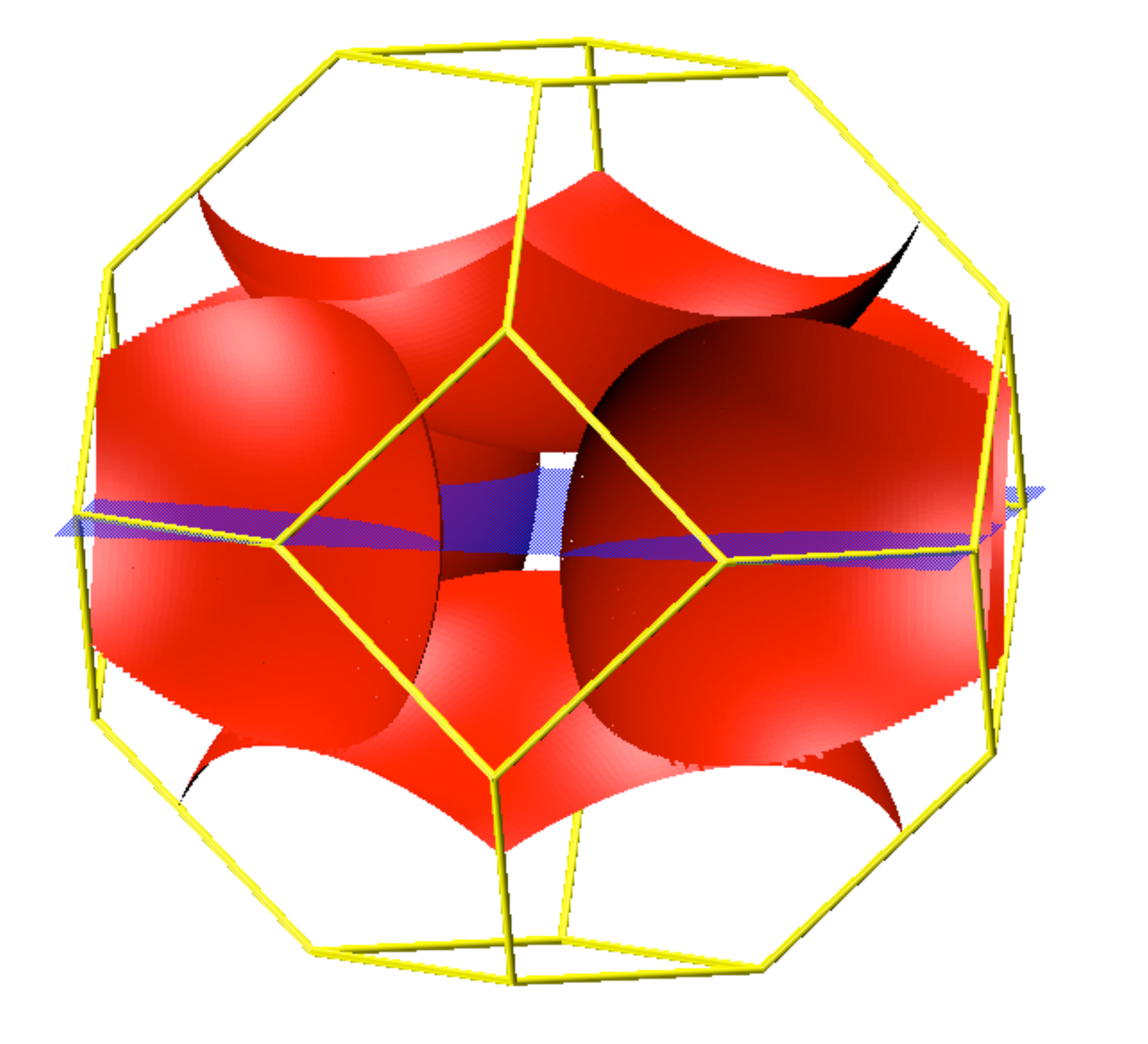}}}
\caption{\label{WS} Interstitial Wigner Seitz cells. The red surfaces represent excluding
spheres at close packing (one sphere is removed in the FCC cell for visibility). Indicated
in blue is the plane in which the density for Figure \ref{plane density} is computed and
the green cylinder in the FCC cell represents the domain for Figure \ref{FCC density
111}.}
\end{figure}

This behavior is shown to follow from a simple but quite different physical model which
focusses on the increasing role played by core exclusion (and near neighbor core
proximity) upon systematic increase of density, at least at energies typified by Fermi
energies. First, consider a system of \textit{N} valence electrons and \textit{N} spheres
(the "ions") occupying a common volume $\Omega$. Though many of the points to be
made are general, the nature of the problem is illustrated with two simple lattices, here
cubic Bravais lattices (BCC and FCC), with sites $\{\textbf R\}$, and the cores will
ultimately be taken to rigorously exclude the valence electrons. As in the formulation of
the NFE model, electrons are also initially taken as non-interacting, though correlation
effects will be of later significance. We consider the thermodynamic limit $N\rightarrow
\infty,\Omega \rightarrow \infty,\Omega/N=4\pi r_s^3a_0^3/3$ and use periodic Von 
K\'arm\'an boundary conditions on $\partial \Omega$. As a starting point, the 
ion-electron interaction $v$ is modeled as (in Hartrees):
\begin{eqnarray}
v(\textbf r)&=&V_0\Theta(r_c-\frac{r}{a_0}).
\end{eqnarray}
where $a_0$ is the Bohr radius, $\Theta$ is the Heaviside function, $r_c a_0$ is the
ionic radius and $V_0$ reflects how strongly excluding the potential is. The periodic
non-interacting electronic Hamiltonian is thus:
\begin{eqnarray}
\label{hamiltonian}
\hat{h}_e(\textbf r)&=&-\frac{\hbar^2}{2m_e}\nabla^2+\sum_{\textbf R}
v(\textbf r-\textbf R),
\end{eqnarray}
for which the Schr\"odinger equation is
\begin{eqnarray}
\label{schrodinger}
\hat{h}_e({\textbf r})\psi_i({\textbf r})&=&\epsilon_i\psi_i({\textbf r})
\end{eqnarray}
(a paramagnetic ground state is again initially assumed). As with the NFE approach
Bloch's theorem fixes the general form of solution of (\ref{schrodinger}). Using scaled
variables:
\begin{eqnarray}
\bar\Omega&=&\frac{\Omega}{(r_s a_0)^3},\hspace{2mm}{\textbf X} = 
\frac{\textbf R}{r_s a_0},\hspace{2mm}{\textbf x} = \frac{\textbf r}{r_s a_0}
\nonumber\\
\bar\nabla &=& r_s a_0\nabla,\hspace{2mm}
\bar{V}_0(r_s)=r_s^2 V_0,\hspace{2mm}\bar{r}_c=\frac{r_c}{r_s},
\end{eqnarray}
the dimensionless form of (\ref{hamiltonian}) is:
\begin{eqnarray}
\bar{h}_e({\textbf x})&=&-\frac{1}{2}\bar\nabla^2+\sum_{\textbf X}
\bar V_0(r_s)\Theta(\bar{r}_c-|{\textbf x-\textbf X}|).
\end{eqnarray}
For fixed $\bar{r}_c=r_c/r_s$, two values of $V_0$ lead to renormalized  eigenvalue
equations independent of $r_s$; $V_0=0$ and $V_0\rightarrow \infty$, the latter case
related to the Quantum Lorentz Gas with a periodic configuration of scatterers
\cite{QLG}. The classical limit will have links to certain billiards problems.

The two-dimensional equivalent yields interesting insight into the problem ahead. Consider
a hexagonal lattice of discs also taken as largely impenetrable to the electrons, as may
be obtained by considering a (111) lattice plane either for BCC or FCC. At the limiting
packing fraction in the plane of $\pi/ 2\sqrt{3}$, the circular regions touch, and per unit
cell there remain two involuted triangular regions either of which must eventually contain
a single electron leading to discrete but highly degenerate and eventually polar ground
states. These triangular regions are essentially isolated and we know the band-width
must vanish at close packing in this case. As the radius of the discs approach their close
packing value, the triangular regions are connected by narrow windows and finite 
band-widths must result. The proximity of excluding discs (and their associated boundary
conditions) will tend to heavily suppress the electronic density at these windows, making
it more favorable energetically for the density to accumulate at the center of a
triangular region.  The band-width is expected to decay rapidly to zero as the radius of
the discs approach their close packing value. Calculations carried out using the method
described next show that this is exactly so. In three dimensions, it is not possible to
fully isolate an interstitial volume, even at close packing; however, we expect the same
argument to be valid, \textit{ie} as the length scale associated with the windows becomes
much smaller than the length scale associated with the inner region of the interstitial
Wigner Seitz (IWS) cell, the band-width should decrease rapidly.

The associated eigenvalue problem can be formulated using a plane wave basis. For set 
$\{\textbf K\}$, the reciprocal lattice, $\textbf k$ a point in the first Brillouin Zone
(1BZ), and $n$ a band index, the eigenvalue problem proceeds from:
\begin{eqnarray}
\label{states}
\braket{\textbf r}{\psi_{n\textbf k}}&\equiv&\frac{1}{\sqrt{\Omega}}\sum_{\textbf K}
e^{i(\textbf{K+k})\cdot \textbf r}c_{n}(\textbf k)_{\textbf K},\\
h_e({\textbf k})_{\textbf K,\textbf K'}&\equiv&\bra{{\textbf{K+k}}}\hat{h}_e
\ket{{\textbf K'+\textbf k}},
\end{eqnarray}
yielding the standard
\begin{eqnarray}
\epsilon_{n\textbf k}c_{n}({\textbf k})_{\textbf K}&=&\sum_{\textbf K'}
h_e({\textbf k})_{\textbf K,\textbf K'}c_{n}({\textbf k})_{\textbf K'}.
\end{eqnarray}
For finite $V_0$, we consider an initial finite plane wave expansion. Reciprocal lattice
vectors $\textbf K$ are included in complete stars, all having a radius smaller than a
cutoff $K_\Lambda$. Sums on the 1BZ are carried out using the special point technique
\cite{ChadiCohen}\cite{RoganLagos}.

The formal limit $V_0\rightarrow \infty$ must eventually be considered as a boundary
condition on the electronic orbitals, by demanding that they vanish on the surfaces of
the excluding spheres. Implementing such a condition completely is quite complex and we
use an approximate alternative: in the limit of large values of $V_0$, as is physically
clear, the results become insensitive to the value of $V_0$ and thus approximate the
limit $V_0\rightarrow \infty$. Convergence can be assessed by computing the lowest
eigenvalue at $\Gamma$ (the center of the 1BZ), at special points on the surface of the
1BZ (H,N,P for the BCC lattice, X,W,L,K for the FCC lattice) and at points
$\textbf k$ pointing towards these special points and of magnitude 
$|{\textbf k}|=k^0_F$ (the free electron Fermi wave vector). The parameters $V_0$
and $K_\Lambda$ were chosen to produce reasonable convergence with moderate
computational effort\cite{methodcomment}.

\begin{figure}
\resizebox{70mm}{!}{\includegraphics{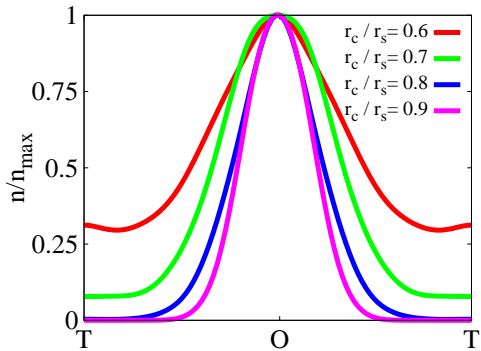}}
\caption{\label{FCC density 111} Electronic density along the main body diagonal of the
IWS cell of the FCC lattice for various values of the ratio $r_c / r_s$. The density is
normalized to its maximum value at O for comparison purposes. T and O are the
tetrahedral and octahedral sites.}
\end{figure}
\begin{figure}
\subfigure[ FCC: $r_c/r_s = 0.5$]{
\resizebox{60mm}{!}{\includegraphics{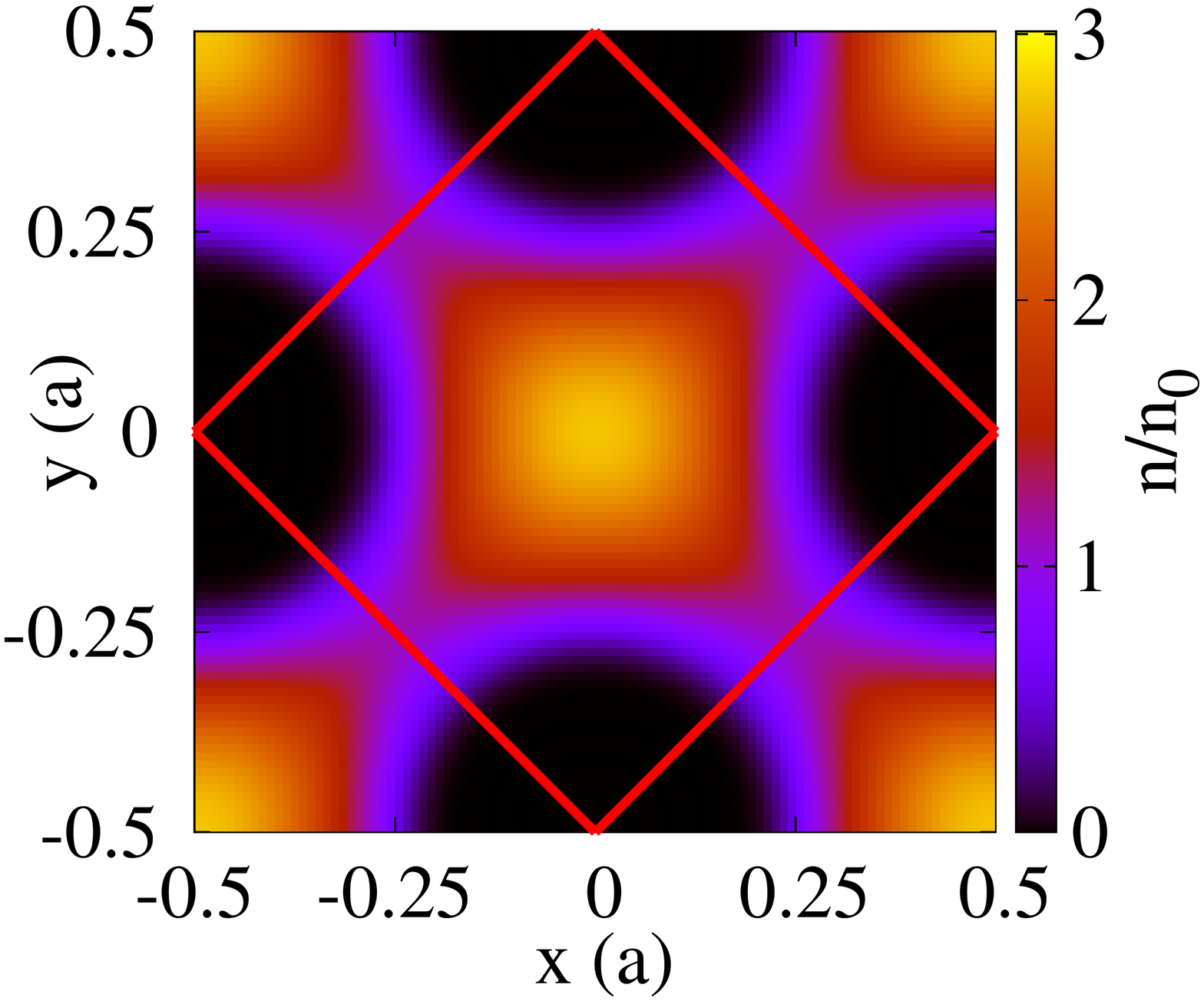}}}
\subfigure[ FCC: $r_c/r_s = 0.7$]{
\resizebox{60mm}{!}{\includegraphics{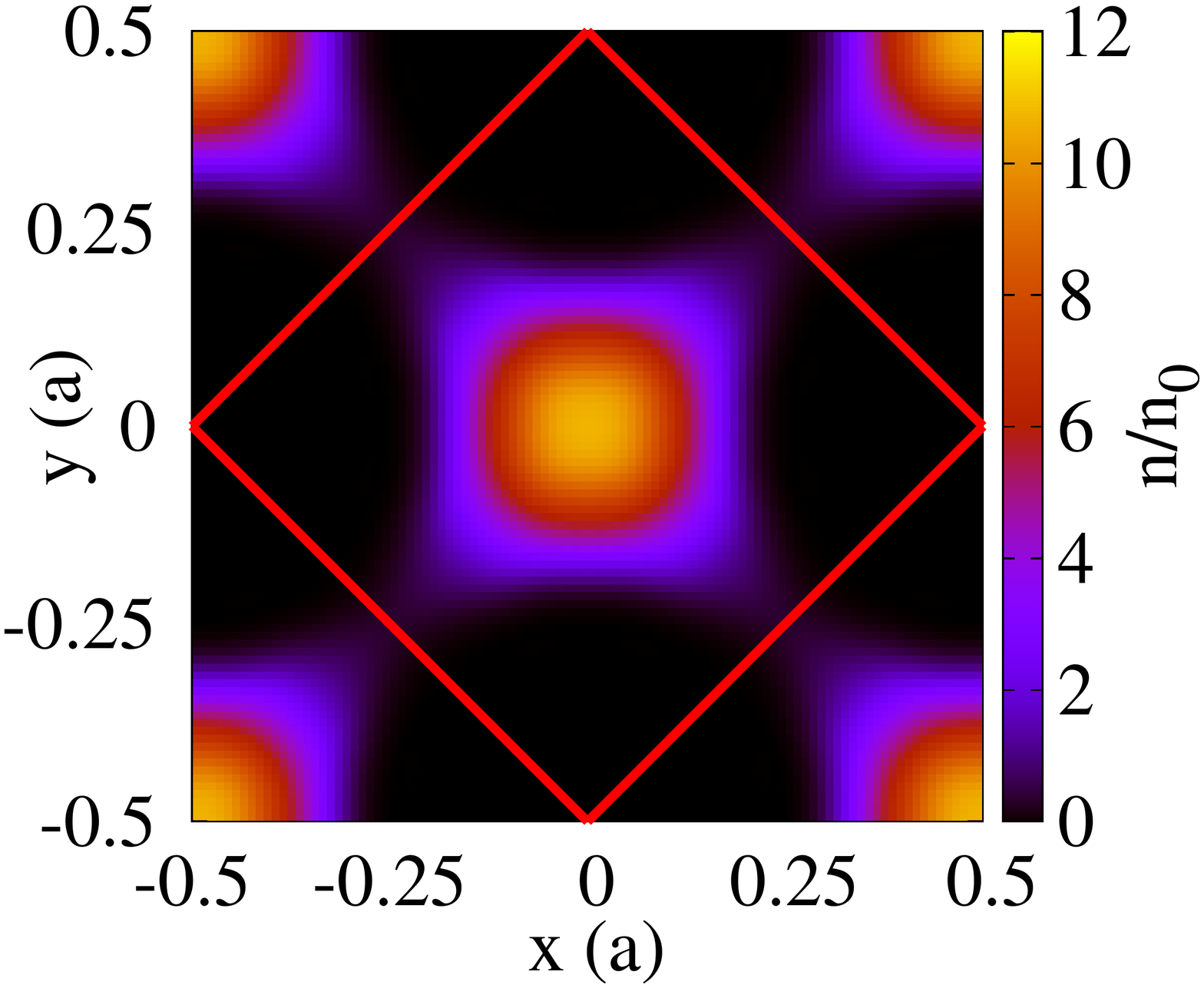}}}
\subfigure[ BCC: $r_c/r_s = 0.5$]{
\resizebox{60mm}{!}{\includegraphics{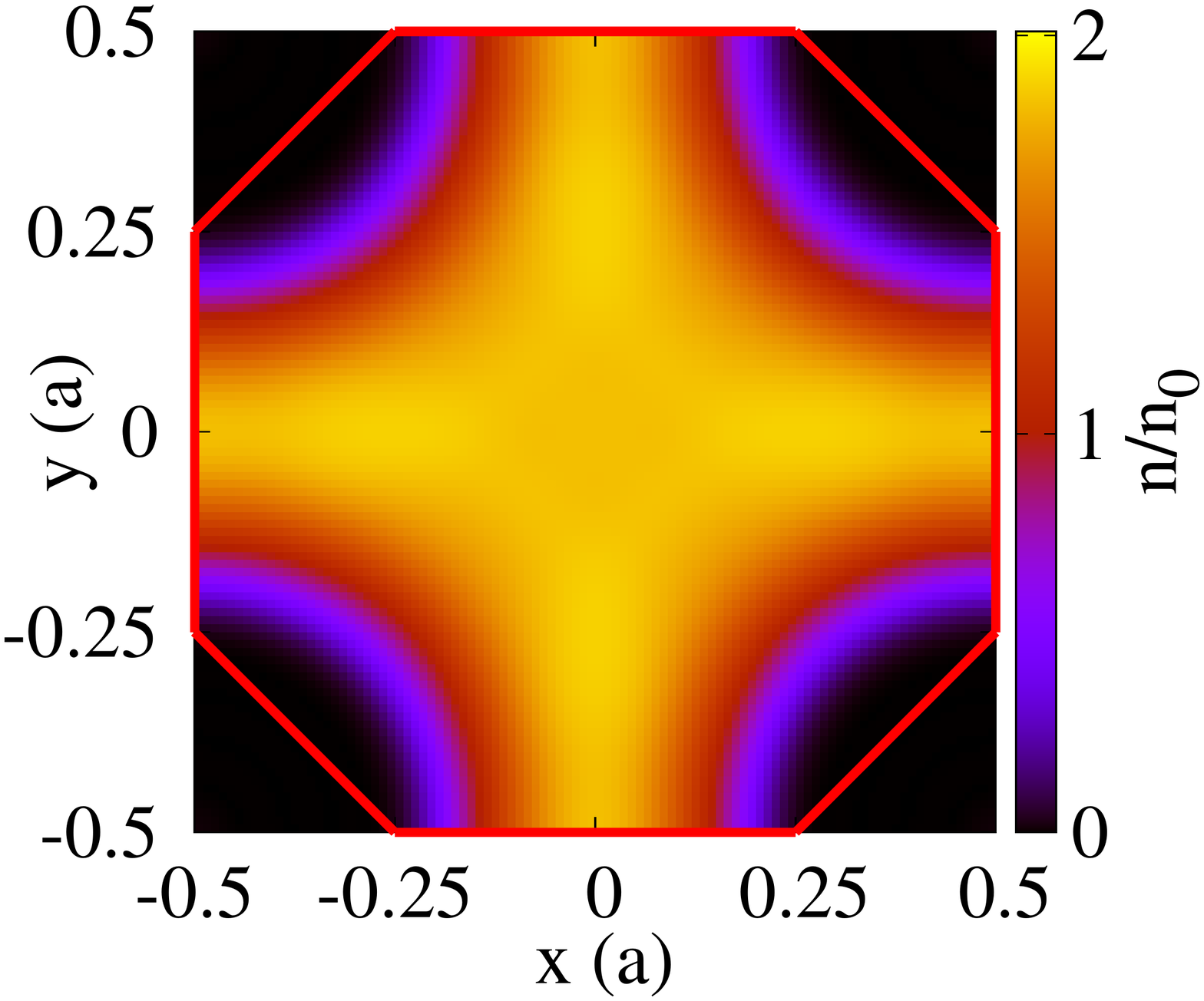}}}
\subfigure[ BCC: $r_c/r_s = 0.7$]{
\resizebox{60mm}{!}{\includegraphics{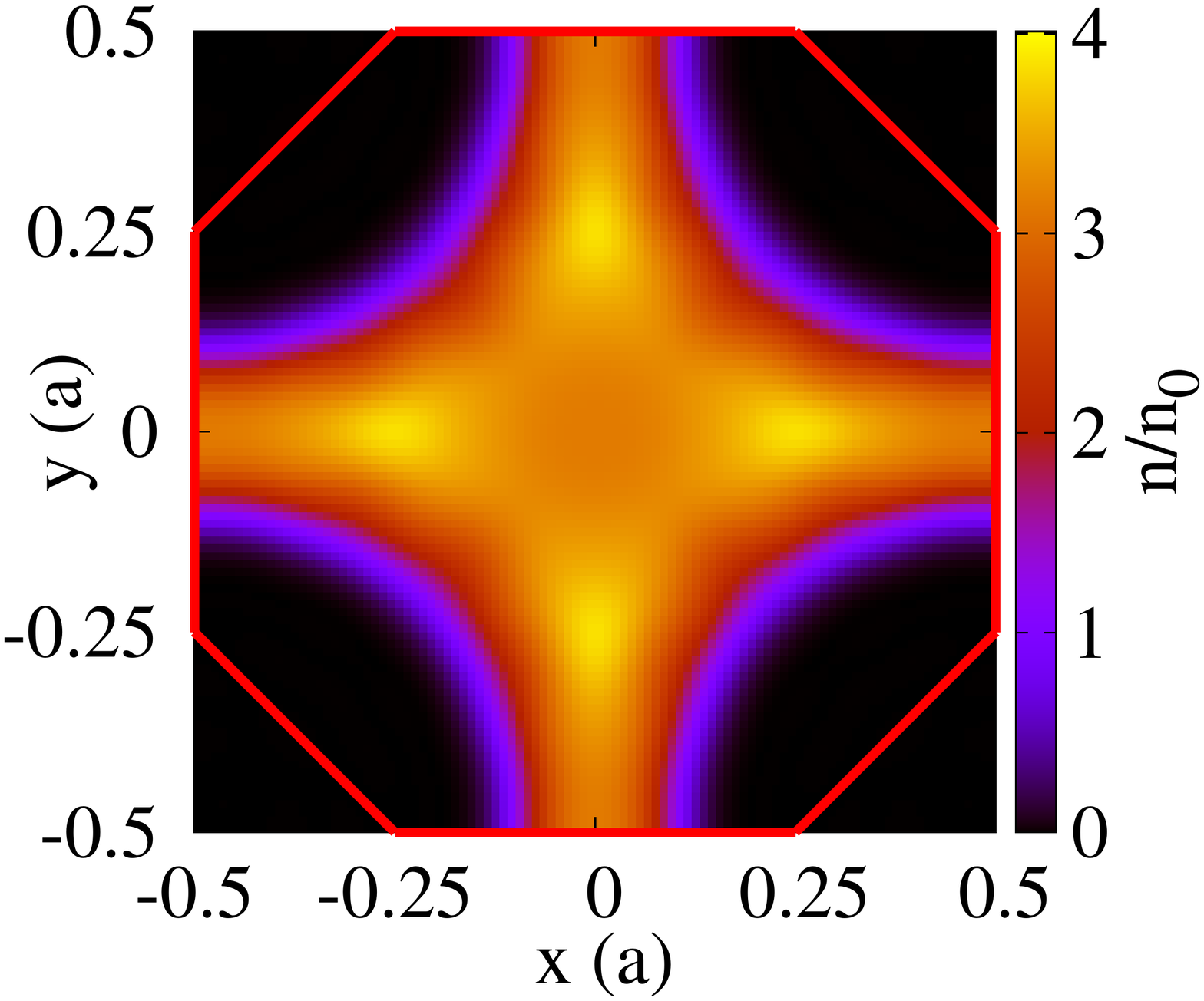}}}
\caption{\label{plane density} Electronic density (from (\ref{schrodinger})) in the z=0
plane. The intersection between the IWS cell and the z=0 plane is outlined in red. Note
that on the approach to close packing, the electronic density develops maxima at
positions in the IWS cell that are maximally distant from the neighboring spheres.}
\end{figure}

From the solution for the states (\ref{states}) we determined the electronic density,
centering the Wigner Seitz cell around an interstitial site (defining an IWS cell, see
Figure \ref{WS}). In the coordinates of the standard cell the octahedral and the
tetrahedral points of the FCC lattice have coordinates 
$(\frac{1}{2},\frac{1}{2},\frac{1}{2})$ and $(\frac{1}{4},\frac{1}{4},\frac{1}{4})$ and
the interstitial site of the BCC lattice has coordinates $(0,0,\frac{1}{2})$ (these choices
are not,of course, unique). The density has been computed on a plane of constant z value
centered on the interstitial sites (Figure \ref{plane density}), and also along the body
diagonal of the FCC IWS cell, from tetrahedral to octahedral to tetrahedral site
(Figure \ref{FCC density 111}). It clearly has a localized character and this provides
insight into the different behavior seen for the two lattices as $r_c/r_s$ approaches
close packing. As spheres take up increasing space in the FCC lattice, the only paths
linking cavities centered at the octahedral sites of neighboring IWS cells must go
through their shared tetrahedral site. As can be seen in Figure \ref{FCC density 111},
the boundary conditions efficiently suppress the density at the tetrahedral site as
$r_c/r_s$ approaches close packing, thus effectively turning the inner surface of the
FCC IWS cell (\textit{ie} the surface of the Wigner Seitz cell outside the spheres) into
a surface where density vanishes. Boundary conditions which impose a vanishing of the
electronic density at the surface of the cell imply flat bands. The density at the surface
of the IWS cell of the BCC lattice, on the other hand, is not as dramatically suppressed
even at close packing, allowing for a significant band-width.

\begin{figure}
\resizebox{90mm}{!}{\includegraphics{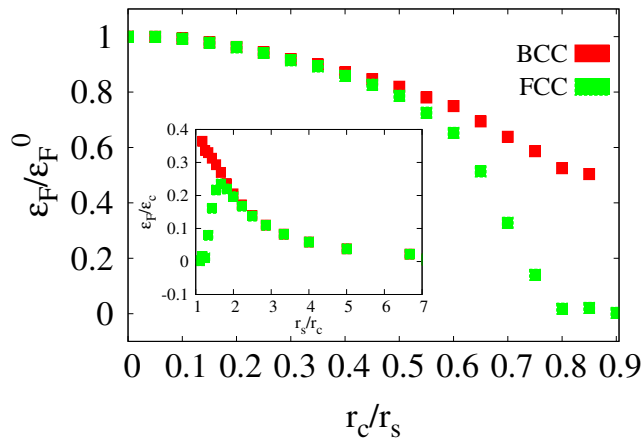}}
\caption{\label{eF} Computed band-width normalized to the free electron Fermi energy as
a function of the ratio $r_c/r_s$. Inset: absolute band-width, normalized to 
$\epsilon_c = \epsilon_F^0(r_s=r_c)$. We observe that the absolute band-width of the
FCC lattice has a turnover around $r_s\simeq 2r_c$, whereas for the BCC case it does
not.}
\end{figure}
\begin{figure}
\subfigure[ FCC: $r_c/r_s = 0.5$]{
\resizebox{60mm}{!}{\includegraphics{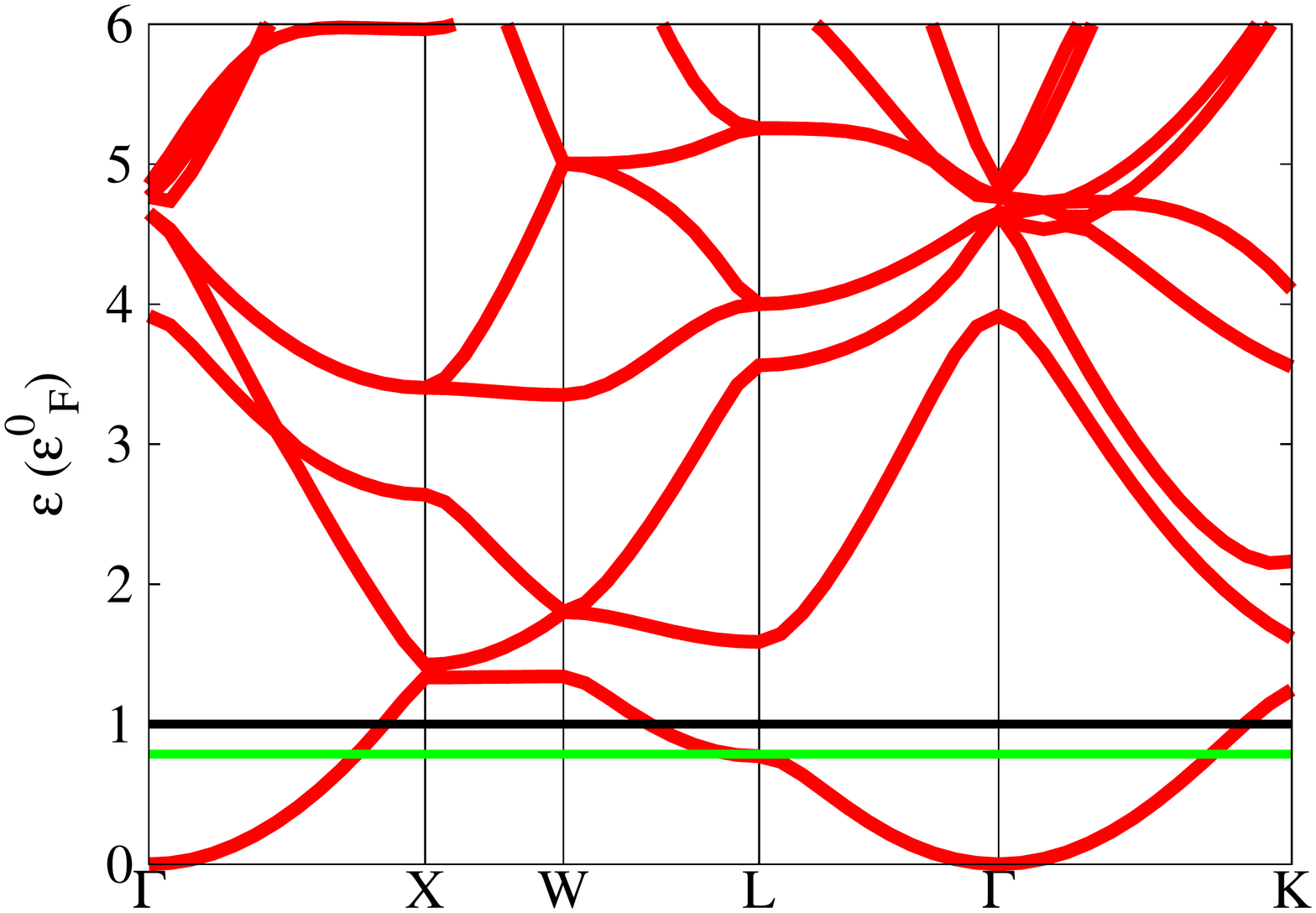}}}
\subfigure[ FCC: $r_c/r_s = 0.7$]{
\resizebox{60mm}{!}{\includegraphics{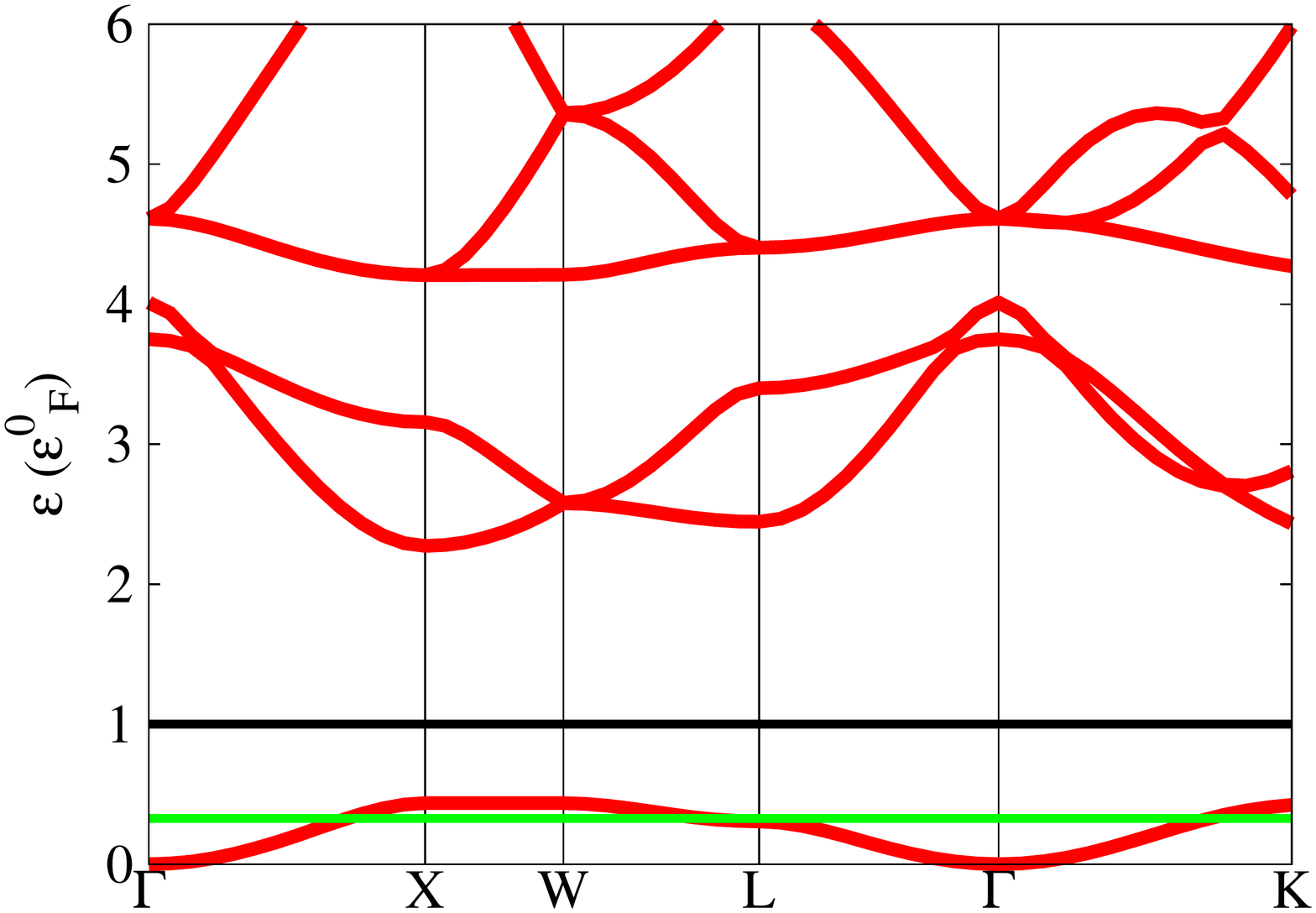}}}
\subfigure[ BCC: $r_c/r_s = 0.5$]{
\resizebox{60mm}{!}{\includegraphics{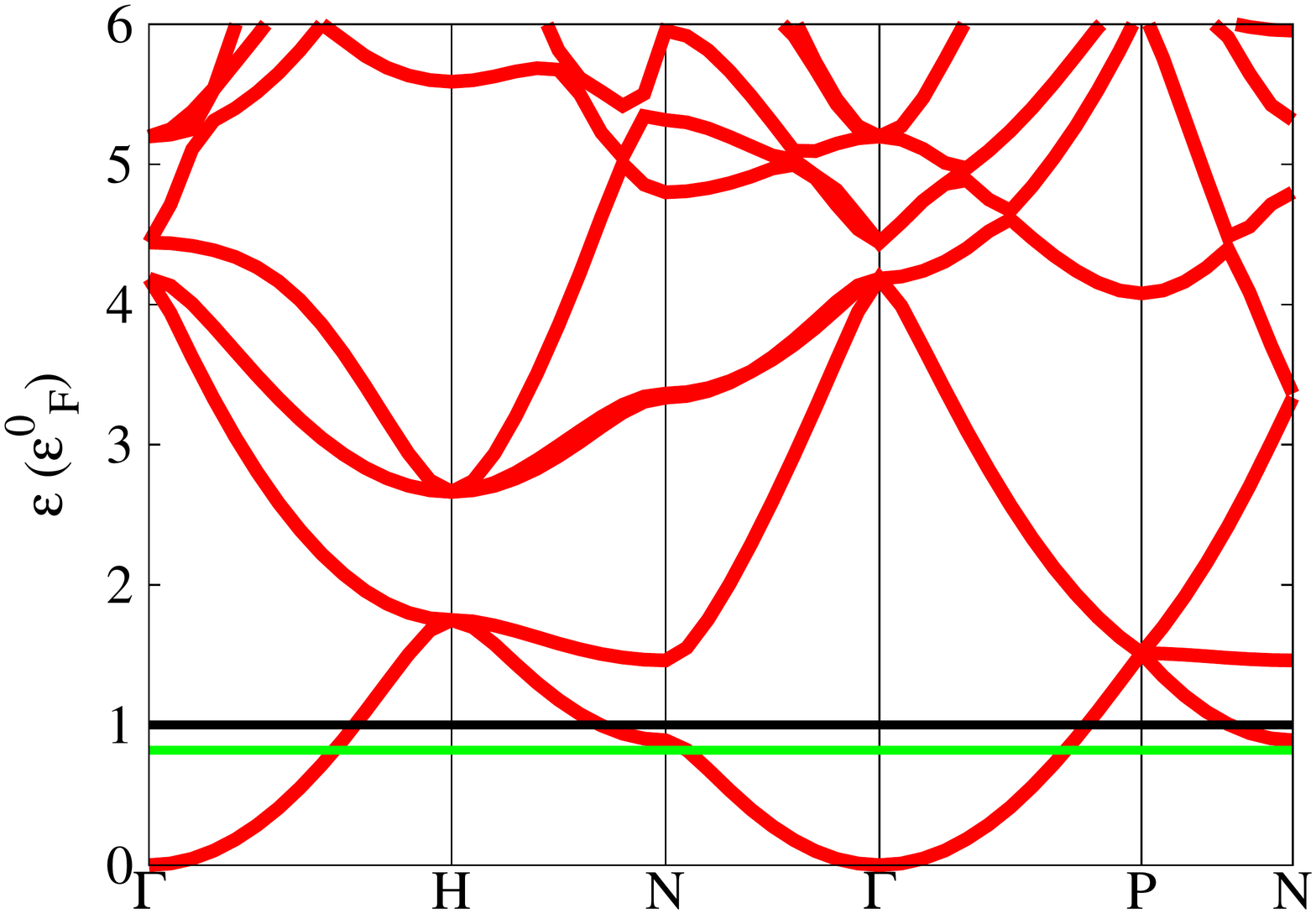}}}
\subfigure[ BCC: $r_c/r_s = 0.7$]{
\resizebox{60mm}{!}{\includegraphics{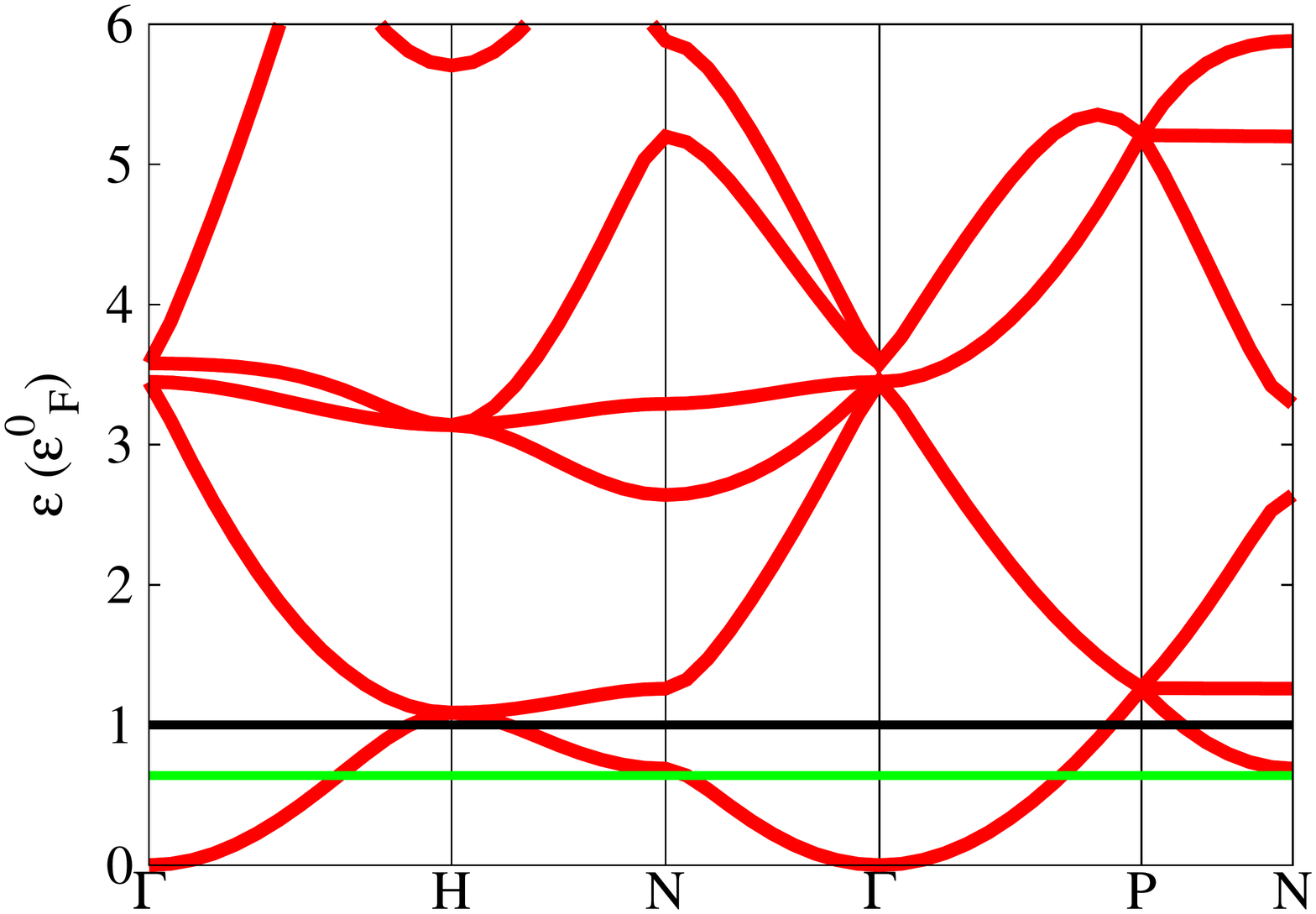}}}
\caption{\label{bands} Eigenvalue spectrum for a few values of $r_c/r_s$ for excluding
spheres in the FCC and BCC lattices. Energies are given in units of $\epsilon_F^0$, the
free electron Fermi energy, and the energy at the bottom of the lowest band is set to
zero. The Fermi energy and the free electron Fermi energy (1 in the graphs' units) are
indicated.}
\end{figure}

Bands and band-widths have been computed in both the FCC and BCC lattices and
representative results can be seen in Figures \ref{eF} and \ref{bands}. Relative 
band-widths decrease monotonically from the free electron value at $r_c/ r_s = 0$
(which we take to yield the empty-lattice bands) to a finite value in the BCC case and to
low values in the FCC case at close packing ($r_c/r_s\simeq 0.905$ for FCC and 
$r_c/r_s\simeq 0.879$ for BCC) . Furthermore, it appears that the Fermi surface
develops necks at the L points of the BCC 1BZ, and almost so at \textit{N} in the FCC
1BZ\cite{neckcomment}.  In order to further substantiate these results, we also
computed the eigenvalues (in units of $\epsilon_F^0$) at $\Gamma$ and at H for the
BCC lattice using an expansion based on Kubic harmonics\cite{Bethe}. We determined the
lowest eigenvalue at $\Gamma$ in the $\alpha$ representation and the lowest
eigenvalues at H in the $\gamma$ and $\alpha$ representations\cite{kubiccomment}. The
difference between the $\gamma$ eigenvalue at H and the $\alpha$ eigenvalue at 
$\Gamma$ decreases for $0 \leq r_c/r_s \leq 0.6$, suggesting strongly that the
relative Fermi energy (the band-width) must follow suit. The method produced 
qualitatively different results according to the number of basis functions used at larger
ratios, suggesting that the Bethe Von de Lage method, which tacitly assumes a 
nuclear-centric view, does not converge quickly as the ratio approaches close packing.

The model simulates core exclusion in the simplest way and proceeds from weak coupling,
NFE like behavior at small relative core volume to narrowing relative band-width (and
narrowing absolute band-width for FCC) at large relative core volume. The rate at which
the bands narrow depends on the connectedness of the IWS cell of the lattice and is
expected therefore to be a general feature. The model reproduces qualitative features
observed by others\cite{Neaton1999}\cite{Neaton2001}, such as the narrowing of the
band-width and strong modulations of the electronic density leading to interstitial site
maxima. It also provides insight into the seemingly strange behavior observed earlier and
is suggested as a possible initiating paradigm for the alkalis under high pressure which
may serve as an alternative to the NFE approach. The ionic cores are not of course
rigorously excluding (as is known from the existence of penetrating states), however the
features of the model should partially subsist so long as the average kinetic energy of
the valence electrons does not exceed the excluding core potential. In that sense, it
provides insight into \textit{ab initio} calculations using valence one pseudopotentials with
repulsive ionic cores. It is unlikely however that the physical system would be well
described by pseudopotentials designed only to reproduce zero pressure properties when
the ionic cores are brought to the point of close packing. 

For the physical situation of interest here it becomes clear that the electrons distribute
themselves in a manner quite different from what the NFE limit would suggest. This can
be deduced by invoking the simplifying features of the Wigner-Seitz approach, but for
the largely confining volumes, themselves initially regarded as spheres. As a first
approximation these have radii $\bar r_s=r_s \{1-(r_c/r_s)^3\}^\frac{1}{3}$ on the
surfaces of which the wave-functions will vanish. The corresponding density is
proportional to $\{\sin \pi (r/a_0\bar r_s) /\pi (r/a_0\bar r_s)\}^2$ and gives, near
close packing, a quite accurate accounting of the density of Figure 
\ref{FCC density 111}, which follows from wave-functions (\ref{states}). The
corresponding scale of kinetic energy, $(\pi/ \bar{r}_s)^2/2$ Ha, immediately suggests
the Hubbard physics character of this problem: as the localization energy cost becomes
large, the energetics of lattice distortions which favor increasing orbital overlap and
reduced band energy may lead to structural phase transitions.

Notice also that the corresponding charge distributions will assume even a partially ionic
form while preserving an essential itinerant or metallic character. Re-introduction of the
Coulomb interactions is not expected to significantly modify the density profile, however
the approach will lead to a quite different viewpoint on the character of the effective
ion-ion interactions. The latter should adopt a somewhat longer ranged form than may be
expected from linear screening arguments in the NFE limit. This has obvious implications
for collective excitations and also for ionic dynamics (the physical character of the liquid
state and melting problem, for example). The behavior of the electronic density also
sheds light on the structural phase transitions which occur at high pressure: the
boundary conditions in the FCC lattice lead to a depletion of the density at the
tetrahedral sites, thus reducing the effective available volume for the electrons beyond
what is already taken away by the impenetrable spheres: a structural phase transition to
a state which avoids such depleted small enclosures and allows the electron liquid to
reduce its zero-point energy by fully occuping available space thus may become
energetically competitive at high pressure. In particular, planar structures allowing the
aggregation of the itinerant charge in effectively two dimensional NFE gases in planar
interstices may become favored. The optical response is also anticipated to be unlike
that of NFE-based systems, though a Drude edge should be preserved. Because of band
narrowing, the density of states at the Fermi level could be high, and the possibility
within a dynamic ionic environment of electron pairing may be enhanced. Some of these
features as well as mixtures and complex structures are presently under examination.
  
\begin{acknowledgments}
This work was supported by the National Science Foundation under grant DMR-0601461.
B. Rousseau would like to acknowledge fruitful discussions with S. Gravel and
Professor S. Bonev.
\end{acknowledgments}

\bibliography{iel_bib}

\begin{thebibliography}{13}
\bibitem[{met()}]{methodcomment}
\bibinfo{note}{The goal of this work is not to produce quantitatively precise
  results but rather to observe the trends in the model and use them as insight
  to understand the behavior of the simple metals under pressure. "Reasonable
  convergence" means that results did not qualitatively change with increase of
  the convergence parameters.}
\bibitem[{nec()}]{neckcomment}
\bibinfo{note}{We note however that it is difficult to resolve whether this
  feature is to be expected for the genuinely impenetrable spheres system or if
  it is an artefact of the finite $V_0$, finite basis size approximation.}
\bibitem[{kub()}]{kubiccomment}
\bibinfo{note}{The boundary conditions were accommodated using an expansion in
  3 and 4 (4 and 6) Kubic harmonics for the $\alpha$ ($\gamma$) representation.
  The boundary conditions where applied as described in\cite{Bethe}.}
\end{thebibliography}
\end{document}